\def\year{2021}\relax
\documentclass[letterpaper]{article} 
\usepackage{aaai21}  
\usepackage{times}  
\usepackage{helvet} 
\usepackage{courier}  
\usepackage[hyphens]{url}  
\usepackage{graphicx} 
\usepackage{natbib}  
\usepackage{caption} 
\usepackage{amssymb}
\usepackage{amsmath}
\usepackage{caption}
\title{Source Separation and Depthwise Separable Convolutions for Computer Audition (Student Abstract)}
\author{

    Gabriel Mersy,
    Jin Hong Kuan \\
}
\affiliations{
    Department of Computer Science and Engineering, University of Minnesota\\
    mersy006@umn.edu, kuan0019@umn.edu
}

\begin{document}

\maketitle

\begin{abstract}
Given recent advances in deep music source separation, we propose a feature representation method that combines source separation with a state-of-the-art representation learning technique that is suitably repurposed for computer audition (i.e. machine listening). We train a depthwise separable convolutional neural network on a challenging electronic dance music (EDM) data set and compare its performance to convolutional neural networks operating on both source separated and standard spectrograms. It is shown that source separation improves classification performance in a limited-data setting compared to the standard single spectrogram approach.

\end{abstract}

\section{Introduction}
Many computer audition methods that map \textit{macro} scale representations of a song---such as a single spectrogram---to music genre taxonomies fail to account for discriminative representations occurring at the \textit{micro} scale---such as instrument rhythm or timbre. This notable limitation can in turn affect the ability of a learner to quickly discern characteristics that are necessary to the classification of similar music genres. To introduce the problem at hand, we conduct a simple empirical experiment on the GTZAN data set which is a traditional benchmark for music classification \cite{gtzan_2002}. 

Consider two sets of music genres $G_1, \, G_2$ where $G_1 = G_2$. We define two non-reflexive relations $g_1 S g_2, \, g_1 D g_2$ on these sets where the former depicts two acoustically \textit{similar} genres, whereas the latter depicts two acoustically \textit{different} genres. Next, two binary classification tasks are constructed to examine performance differences between these two relations for an arbitrary convolutional neural network. It is evident that the model quickly reaches an accuracy of 100\% on the task with heavy metal and jazz (difference relation $D$) but struggles to achieve comparable performance on the task with rock and country (accuracy of 27\%; similarity relation $S$). With few data and little time, how might a learner compute salient features to discriminate between alike genres? 

\section{Related Work}
Gjerdingen and Perrott show that human classification of music genre tends to occur within a quarter of a second and that listeners are able to distinguish one genre from another based on a decoding of the song components \cite{doi:10.1080/09298210802479268}. Existing genre classification methods by means of source separation tend to focus on a limited number of sources (e.g. percussive and harmonic) and employ unsupervised techniques such as non-negative matrix factorization to separate the tracks \cite{10.1007/s10844-017-0464-5}. 
\section{Methods}
\textbf{Data set and music segmentation} \;  Five genres from the the GiantSteps+ EDM Key Data set were included, and these genres were combined with an additional genre from a set of free download songs\footnote{u/Futureops 2017; Free download collection... \textit{Reddit}.} for a total of \textit{66 songs} spanning \textit{six genres} \cite{Knees2015TwoDS}. These genres are all closely related which by the logic of the introductory GTZAN experiment should present a challenging classification task. A human interactive technique was employed to segment each song. The length of each audio segment was then trimmed to approximately 10 seconds at a sample rate of 22,050 hertz.   

\noindent \textbf{Feature representation} \; To obtain the feature representation, a U-net model was used to divide each stereo audio signal $\mathbf{X}(t)$ into its musical constituents such that
\begin{equation}
    \mathbf{X}(t) = \mathbf{\tilde{S}}_{b}(t) + \mathbf{\tilde{S}}_{d}(t) + \mathbf{\tilde{S}}_{o}(t) + \mathbf{\tilde{S}}_{v}(t)
\end{equation}
\noindent where $\mathbf{\tilde{S}}_i(t) \in I$ represents the estimated audio sources in the set of instruments $I$: bass, drums, other, and vocals respectively \cite{Hennequin2019SPLEETERAF}. Each stereo instrument $\mathbf{\tilde{S}}_i(t)$ in (1) was replaced with a mono mel spectrogram $\mathbf{\tilde{M}}_i \in I$ that captures the spectral frequency response. By stacking each spectrogram depthwise, a feature representation that is analogous to a color image was derived---where each channel is composed of an instrument source spectrogram corresponding to one of three instrument types (vocals were not included). The feature tensor for the source separated representation has the shape $(66, 128, 458, 3)$. Two feature tensors with the shape $(66, 128, 458, 1)$ were computed for the purpose of comparison to baseline methods---only one of which included vocals.  

\noindent \textbf{Neural networks} \; The four neural networks were composed of two convolution layers each followed by a max pooling layer. Class weights were added to attenuate the unbalanced nature of the data set. Two of these models operated on the source separated tensor: standard 2D multichannel convolution and 2D depthwise separable convolution. One standard single-spectrogram network operated on the tensor that includes vocals, while the other network operated on the tensor that does not include vocals. 

In contrast to multichannel 2D convolution which convolves a 3D filter $\mathcal{W} \in \mathbb{R}^3$ over the image channels of $\mathcal{Y}$, 2D depthwise separable convolution applies an identical filter $\mathbf{W}_d \in \mathbb{R}^2$ to each of the $c$ channels separately to produce $c$ feature maps. As seen in (2), these feature maps are then combined with a pointwise $1 \times 1 \times c$ convolution filter $\mathcal{W}_p$ \cite{Kaiser2018DepthwiseSC}.
\begin{equation}
    \begin{split}
    \mathrm{SepConv}(\mathcal{W}_p, \mathbf{W}_d, \mathcal{Y}) = \mathrm{PointwiseConv_{(i,j)}}(\mathcal{W}_p,\\\mathrm{DepthwiseConv}_{(i,j)}(\mathbf{W}_d, \mathcal{Y}))
\end{split}
\end{equation}
\noindent With $c = 3$ channels and a receptive field size of $k$, for any $k > 1$ depthwise separable convolution reduces the number of parameters per convolution from $9k$ to $3k + 9$ thereby proportionately reducing computation time.
Given that a small sample size generally results in parameter gradients that have high variance, architectural techniques were implemented to increase model bias. Network parameters were regularized with penalties on both the L1 and L2 norms, batch normalization was applied after each convolution layer, and a dropout layer with $p = 0.5$ was introduced immediately prior to the fully connected layer.

\noindent \textbf{Model evaluation} To improve the validity of the experiments, we introduce two metrics that are based on the sampling distributions of model accuracy and F1 across multiple trials. For each independent trial, 80\% train-20\% test splits were carried out at a unique random seed to produce identical train and tests sets for each model. The models were then trained for six epochs using sparse categorical crossentropy loss, and metrics were recorded for the epoch with the minimum test loss. The mean for each metric after 50 trials is reported in Table 1. The F1 distributions for the 3 and 1 spectrogram models after 50 trials are presented in Figure 1.  

\begin{table}[h]
\centering
\begin{tabular}{ |c|c|c|c| } 
 \hline
 \textbf{Model} & \textbf{Accuracy} & \textbf{F1} \\ 
 \hline
1 spec, 2DConv-Full & 0.380 &  0.350 \\ 
 \hline
1 spec, 2DConv-No-Vox\dag & 0.384 & 0.371 \\
 \hline
 3 spec, 2DConv & \textbf{0.456} & \textbf{0.435} \\ 
 \hline
 3 spec, DW-2DConv & 0.451 & 0.419 \\
 \hline
\end{tabular}
\caption{\label{tab:results} Mean test set metrics for 50 trials; \dag baseline}
\end{table}

\begin{figure}
    \centering
    \includegraphics[scale=0.04]{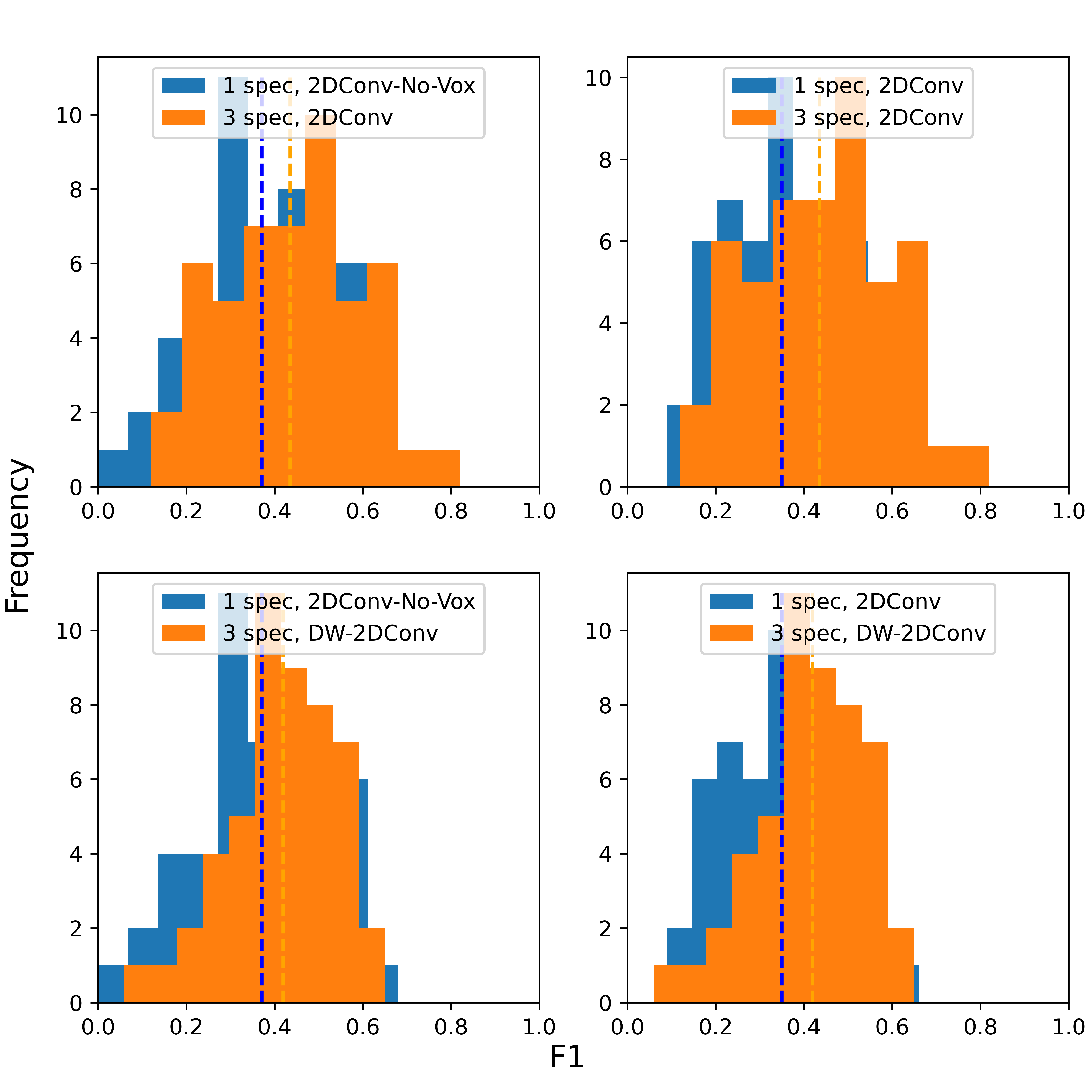}
    \caption{F1 scores of 3 spectrogram (orange) and 1 spectrogram (blue) models; dashed line denotes the mean}
    \label{fig:my_label}
\end{figure}

\section{Conclusion}
Each of the source separated 3 spectrogram models significantly outperformed the 1 spectrogram baseline. The 3 spectrogram 2D convolution model marginally outperformed the depthwise convolution model when each was compared to the baseline (F1 p-values of $p = 0.04$ and $p = 0.07$ respectively). The underperformance of the depthwise model is best understood by its left-skewed F1 distribution, which is markedly different from the Gaussian F1 distribution of the 2D convolution model. This instability could be elucidated by the depthwise model's reduced parameter quantity. It is also observed that the inclusion of vocals had little effect on classifier performance in this setting. Ultimately, two songs from the same genre presumably have paired instrument channels that have similar temporal and spectral patterns, while this may not be case for two songs from different genres. Thus, we believe that the improved performance of the 3 spectrogram models can be attributed to information gained from the introduction of source separation methods. 



\bibliography{bibliography.bib}

\end{document}